\documentclass{article}
\usepackage{arxiv}

\usepackage[utf8]{inputenc} 
\usepackage[T1]{fontenc}    
\usepackage{amsfonts}       

\usepackage{adjustbox}

\usepackage{stackengine}

\usepackage{hyperref} 
\usepackage{doi} 

\usepackage{cleveref}       
\usepackage{natbib}

\author{Ingmar Böschen}
\title{\emph{statcheck} is flawed by design and no valid spell\\checker for statistical results}

\date{}

\newif\ifuniqueAffiliation
\uniqueAffiliationtrue

\ifuniqueAffiliation 
\author{ \href{https://orcid.org/0000-0003-1159-3991}{\includegraphics[scale=0.06]{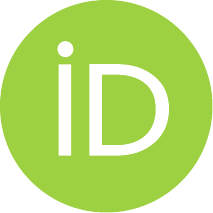}\hspace{1mm}Ingmar~Böschen}\\
    Institute of Psychology\\	
	Research Methods and Statistics\\
	University of Hamburg\\
	\texttt{ingmar.boeschen@uni-hamburg.de} \\
}
\else
\usepackage{authblk}

\newbox{\orcid}\sbox{\orcid}{\includegraphics[scale=0.06]{orcid.pdf}} 
\author[1]{%
	\href{https://orcid.org/0000-0003-1159-3991}{\usebox{\orcid}\hspace{1mm}Ingmar~Boeschen\thanks{\texttt{ingmar.boeschen@uni-hamburg.de}}}%
}

\affil[1]{Department of Research Methods and Statistics, University of Hamburg, Hamburg, Germany}
\fi

\renewcommand{\headeright}{}
\renewcommand{\undertitle}{Technical Report}
\renewcommand{\shorttitle}{\textit{statcheck} is flawed by design and no valid spell checker for statistical results}

\hypersetup{
pdftitle={statcheck is flawed by design and no valid spell checker for statistical results},
pdfsubject={stat.AP, stat.CO},
pdfauthor={Ingmar~Böschen},
pdfkeywords={software, feature extraction, statcheck},
}

\begin{document}
\maketitle

\begin{abstract}
The R package \textit{statcheck} is designed to extract statistical test results from text and check the consistency of the reported test statistics and corresponding p-values. 
Recently, it has also been featured as a spell checker for statistical results, aimed at improving reporting accuracy in scientific publications. 

In this study, I perform a check on \textit{statcheck} using a non-exhaustive list of 187 simple text strings with arbitrary statistical test results. These strings represent a wide range of textual representations of results including correctly manageable results, non-targeted test statistics, variable reporting styles, and common typos.

Since \textit{statcheck}'s detection heuristic is tied to a specific set of statistical test results that strictly adhere to the American Psychological Association (APA) reporting guidelines, it is unable to detect and check any reported result that even slightly deviates from this narrow style. 
In practice, \textit{statcheck} is unlikely to detect many statistical test results reported in the literature. 

I conclude that the capabilities and usefulness of the \textit{statcheck} software are very limited and that it should not be used to detect irregularities in results nor as a spell checker for statistical results. 

Future developments should aim to incorporate more flexible algorithms capable of handling a broader variety of reporting styles, such as those provided by \textit{JATSdecoder} and Large Language Models, which show promise in overcoming these limitations but they cannot replace the critical eye of a knowledgeable reader. 

\end{abstract}

\section{Objective}
\label{sec:intro}

Spell-checking software, a standard feature in most text processing tools, aids in avoiding typographical errors.  Similarly, the R \citep{RCore} package \textit{statcheck} \citep{statcheck1.4} was developed to enhance the integrity of statistical test result reporting by checking the consistency of test statistics and corresponding p-values in textual ressources. 
\textit{statcheck} has been introduced as a helpful assistant to identify inconsistencies in scientific reports, catering to authors, reviewers and journal quality management systems  alike \citep{baker2016stat, singh2017controversial, nuijten2023implementing}. 
Recently, \textit{statcheck} has been featured as a spell checker for statistical results \citep{statcheckSpellChecker}. 
In general, it is crucial for error detection and spell-checking software to be reliable. 
Spellcheckers ought to alert users to potential errors and provide suggestions for corrections for all possible error types. 

I completely agree with the authors of \textit{statcheck} that the consistent and complete reporting of statistical results in scientific reports is essential for credibility. 
Moreover, an automated extraction of results can bring benefits for meta-analyses and enable study identification by effect size and sample size. 

To investigate the extent of the use of nil-null hypothesis significance testing and the rate of significant results, I employed \textit{statcheck} to extract the statistical results from a vast corpus of articles by multiple distributors. 
It turned out that \textit{statcheck} is an inappropriate tool to perform the intended analysis on the prevalence of statistical results in the scientific literature, nor that it can serve as a reliable results spell checker. 

As Schmidt already pointed out \citep{schmidt2017statcheck, schmidt2016sources}, \textit{statcheck} suffers several severe limitations and drawbacks as an error detector. 
Schmidt indicated that \textit{statcheck} is only capable of detecting and checking results that are reported in a way that strictly follows the guidelines of the American Psychological Association (APA). 
The general ability of \textit{statcheck} to extract reports of statistical results from text is weakened by these semantic and further technical restrictions (letter encoding, indexing). 
Further, Schmidt \citep{schmidt2017statcheck} reused the results of the evaluation study by Nuijten et al. \citep{nuijten2017validity} and attested \textit{statcheck} poor sensitivity ($.52$) and poor validity ($\phi = .54$).

Also, the latest \textit{statcheck} update in 2023 \citep{statcheck1.4} does not contain fixes for any of the numerous bugs that I pointed out earlier \citep{getStats} when I introduced the \textit{get.stats} function \citep{getStats} out of the \textit{JATSdecoder} \citep{JATSdecoderCRAN1.2} package.
Compared to \textit{statcheck}, \textit{get.stats} is a general and robust tool for extracting numerical results from text. 
Unlike \textit{statcheck}, \textit{get.stats} extracts results without any restrictions on reporting style. 
Although \textit{get.stats} is capable of identifying and processing misspelled and inconsistent results, it should be noted that it is not designed to be a spell checker for results. 

The checks on \textit{statcheck} presented here start with some examples of results that \textit{statcheck} can reliably extract and check. 
Then, a list of reasonable results reported in APA style is showcased that \textit{statcheck} cannot handle properly. 
Ultimately, the focus is on the most problematic property of \textit{statcheck}, the high non-detection rate of individual reporting styles and typos. 
A manually curated, non-exhaustive list with general result patterns is presented, none of which \textit{statcheck} detects and therefore cannot check. 

\section{Possible causes for inconsistencies in the presentation of results}
Scientific articles usually contain many numerical results, which often contain relevant decimal places and therefore have many characters.
Results are often manually transcribed from the output of statistical software solutions into the text of a report, which in practice may result in transcription errors. 
Numerical transmission errors are likely to represent the most common source of inconsistencies in reports of statistical results. 
Manual transcription errors may result in numbers being substituted, omitted, or added to the actual results. 
It is also conceivable that the test statistic and the p-value of two test results are swapped. 
In addition, incorrect rounding can lead to inconsistencies. 
In the worst case, a transcription or rounding error leads to the wrong decision about the tested hypothesis and, therefore, to false inferences. 
It would be very beneficial to avoid detectable and obviously inconsistently reported results, and I admit that \textit{statcheck} is a first attempt to help here. 

However, there are also plausible reasons why a reported p-value might be mislabeled as inconsistent with the test statistic, such as the use of directional tests or a correction for multiple testing. 
Some statistical tests, such as the t-test, which is also the standard in multiple regression analyses, permit the testing of directed or undirected hypotheses. 
The p-value obtained from a one-tailed test and the resulting inference may differ substantially from those of a two-tailed calculation. 
Furthermore, the p-value of a directional test can take either $p/2$ or $1-p/2$ (where p denotes the p-value of a two-tailed test), depending on the direction of the tested hypothesis. 

Statistical test results can be corrected for multiple testing to compensate for $\alpha$ error accumulation. 
This correction can be made either by using a corrected $\alpha$ level for decision-making or by correcting the resulting p-values and using the origial $\alpha$. 
Comparing a corrected and an uncorrected p-value can lead to false-positive detections of inconsistencies if the numerical results are significantly different. 
A major problem with any automated result checking is that it probably does not take into account that the alpha-level or p-values should be corrected if several tests where performed on the same data, to preserve the original $\alpha$. 
The problematic consequence is that uncorrected test results are confirmed as consistent when a correction should have been applied, and the result is not critically analyzed further.

\section{Some consistency checks on \textit{statcheck}}
\subsection{Results that \textit{statcheck} detects and checks}
Before discussing \textit{statcheck}'s limitations, let's examine the scope of results this tool can process properly. 
Table \ref{tab::statcheckWorks} presents some arbitrary examples of textual representations of statistical test results in APA format, accompanied by the corresponding \textit{statcheck} output. 
All results are correctly extracted and checked for consistency of p-values. 
It is worth noting that every input begins with a space, since \textit{statcheck} cannot recognize Z-statistics that start at the first character of an input string. 

\begin{table}[!ht]
\caption{Example text representations of statistical test results that are checkable with \textit{statcheck} in rearranged output table}
\label{tab::statcheckWorks}
\centering
\scriptsize
\setlength{\tabcolsep}{4pt}
\begin{tabular}{lllllllllllll}
  \hline
No & raw input & extracted by \textit{statcheck}& test\_type & df1 & df2 & test\_comp & test\_value & p\_comp & reported\_p & computed\_p & error & decision\_error \\ 
  \hline
1 & " t(12)=2.3, p$<$.05"  & t(12)=2.3, p$<$.05  & t  & NA  & 12  & =  & 2.30  & $<$  & 0.05  & 0.04019757  & FALSE  & FALSE  \\ 
  2 & " F(1,23)=4.5, p=.23"  & F(1,23)=4.5, p=.23  & F  &  1  & 23  & =  & 4.50  & =  & 0.23  & 0.04488686  & TRUE  & TRUE  \\ 
  3 & " r(12)=.34, p=.56"  & r(12)=.34, p=.56  & r  & NA  & 12  & =  & 0.34  & =  & 0.56  & 0.23428054  & TRUE  & FALSE  \\ 
  4 & " Z=1.2, p$<$.34"  & Z=1.2, p$<$.34  & Z  & NA  & NA  & =  & 1.20  & $<$  & 0.34  & 0.23013934  & FALSE  & FALSE  \\ 
  5 & " $\chi$\verb|^|2(12)=3.4, p$<$.05"  & \verb|^|2(12)=3.4, p$<$.05  & Chi2  & 12  & NA  & =  & 3.40  & $<$  & 0.05  & 0.99200057  & TRUE  & TRUE  \\ 
  6 & " $\chi$2(12)=3.4, p$<$.05"  & $\chi$2(12)=3.4, p$<$.05  & Chi2  & 12  & NA  & =  & 3.40  & $<$  & 0.05  & 0.99200057  & TRUE  & TRUE  \\ 
  7 & " Chi\verb|^|2(12)=3.4, p$<$.05" & \verb|^|2(12)=3.4, p$<$.05  & Chi2  & 12  & NA  & =  & 3.40  & $<$  & 0.05  & 0.99200057  & TRUE  & TRUE  \\ 
  8 & " chi2(12)=3.4, p$<$.05"  & i2(12)=3.4, p$<$.05  & Chi2  & 12  & NA  & =  & 3.40  & $<$  & 0.05  & 0.99200057  & TRUE  & TRUE  \\ 
  9 & " Q(12)=3.4, p$<$.01"  & Q(12)=3.4, p$<$.01  & Q  & 12  & NA  & =  & 3.40  & $<$  & 0.01  & 0.99200057  & TRUE  & TRUE  \\ 
 10 & " t(12)=.34, n.s."  & t(12)=.34, n.s.  & t  & NA  & 12  & =  & 0.34  & ns  & NA  & 0.73973384  & FALSE  & FALSE  \\
\hline
\end{tabular}
\end{table}

The chi-square statistic is listed multiple times in the table, necessitating a more detailed explanation and leading to the next block of checks. 
Within the targeted space of test statistics, it is the only one that is denoted with two special characters, the Greek letter $\chi$ and the superscripted 2. 
Special characters highly complicate the detection of results from text, especially when the text is extracted from PDF files. 
In general, PDF to text conversion may result in a total loss of special characters or cause strange compilation errors (e.g., the number 5 for the operator `=').  
There is considerable technical complexity arising from the fact that various character representations can refer to the same visual depiction of an on-screen symbol. 
This includes both character coding systems (e.g., utf-8, hexadecimal, HTML) and style format (e.g., italic, bold). 
Although this topic will not be explored here, it becomes obvious that the authors of the \textit{statcheck} algorithm bypassed this challenge with a flawed approach.

The following test sets demonstrate that the extraction of chi-square results by \textit{statcheck} is imprecise. 
This is probably the result of the lack of a solution for dealing with the difficulties associated with the Greek letter $\chi$ and the superscript digit 2. 
To better understand the inconsistencies in the extraction of chi-square results, it is worthwhile examining the raw results extracted by \textit{statcheck} (examples 5--8).
There are three distinct ways in which the raw result is returned, with `\textasciicircum2' although it has the Greek letter $\chi$ in front (5), exactly like the input (6), without the spelled-out `Chi' behind `textasciicircum2' (7), and as  `i2' in the case of a spelled-out `chi' followed by the number two (8).

\subsection{Results that \textit{statcheck} mis-identifies as $\chi^2$-statistics}
To test how \textit{statcheck} processes any plausible test statistic, first a set of 52 arbitrary results labeled with every upper and lowercase letter is combined with a number in brackets, a numerical result and a p-value (e.g.: `A(12)=.3, p<.05') and processed with \textit{statcheck}. 
Table \ref{tab:simpleLetters} displays the data frame that \textit{statcheck} returned. 
It consists of 46 detected results, six results (D, F, W, Z, n, z) were neither detected nor checked. 
Only three (the Q, r and t statistics) out of the 46 identified results were treated correctly. 
The remaining 43 non-standard results were erroneously treated as a chi-square statistic. 

\begin{table}[!ht]
\setlength{\tabcolsep}{4pt}
\caption{Truncated output of: \textit{statcheck(paste0(" ",c(LETTERS,letters),"(12)=.3, p<.05"))}}
\label{tab:simpleLetters}

\centering
\scriptsize
\begin{tabular}{rllllllllllllll}
\hline
 & raw & source & test\_type & df1 & df2 & test\_comp & test\_value & p\_comp & reported\_p & computed\_p & error & decision\_error & one\_tailed\_in\_txt \\
  \hline
1 & A(12)=.3, p$<$.05 & 01  & Chi2  & 12  & NA  & =  & 0.2  & $<$  & 0.05  & 1.0000  & TRUE  & TRUE  & FALSE  \\
  2 & B(12)=.3, p$<$.05 & 02  & Chi2  & 12  & NA  & =  & 0.2  & $<$  & 0.05  & 1.0000  & TRUE  & TRUE  & FALSE  \\
  3 & C(12)=.3, p$<$.05 & 03  & Chi2  & 12  & NA  & =  & 0.2  & $<$  & 0.05  & 1.0000  & TRUE  & TRUE  & FALSE  \\
  4 & E(12)=.3, p$<$.05 & 05  & Chi2  & 12  & NA  & =  & 0.2  & $<$  & 0.05  & 1.0000  & TRUE  & TRUE  & FALSE  \\
  5 & G(12)=.3, p$<$.05 & 07  & Chi2  & 12  & NA  & =  & 0.2  & $<$  & 0.05  & 1.0000  & TRUE  & TRUE  & FALSE  \\
...\\
  20 & V(12)=.3, p$<$.05 & 22  & Chi2  & 12  & NA  & =  & 0.2  & $<$  & 0.05  & 1.0000  & TRUE  & TRUE  & FALSE  \\
  21 & X(12)=.3, p$<$.05 & 24  & Chi2  & 12  & NA  & =  & 0.2  & $<$  & 0.05  & 1.0000  & TRUE  & TRUE  & FALSE  \\
  22 & Y(12)=.3, p$<$.05 & 25  & Chi2  & 12  & NA  & =  & 0.2  & $<$  & 0.05  & 1.0000  & TRUE  & TRUE  & FALSE  \\
  23 & a(12)=.3, p$<$.05 & 27  & Chi2  & 12  & NA  & =  & 0.2  & $<$  & 0.05  & 1.0000  & TRUE  & TRUE  & FALSE  \\
...\\
 35 & m(12)=.3, p$<$.05 & 39  & Chi2  & 12  & NA  & =  & 0.3  & $<$  & 0.05  & 1.0000  & TRUE  & TRUE  & FALSE  \\
  36 & o(12)=.3, p$<$.05 & 41  & Chi2  & 12  & NA  & =  & 0.3  & $<$  & 0.05  & 1.0000  & TRUE  & TRUE  & FALSE  \\
...\\
  39 & r(12)=.3, p$<$.05 & 44  & r  & NA  & 12  & =  & 0.2  & $<$  & 0.05  & 0.4930037  & TRUE  & TRUE  & FALSE  \\
  40 & s(12)=.3, p$<$.05 & 45  & Chi2  & 12  & NA  & =  & 0.2  & $<$  & 0.05  & 1.0000  & TRUE  & TRUE  & FALSE  \\
  41 & t(12)=.3, p$<$.05 & 46  & t  & NA  & 12  & =  & 0.2  & $<$  & 0.05  & 0.8448298  & TRUE  & TRUE  & FALSE  \\
  42 & u(12)=.3, p$<$.05 & 47  & Chi2  & 12  & NA  & =  & 0.2  & $<$  & 0.05  & 1.0000  & TRUE  & TRUE  & FALSE  \\
...\\
  46 & y(12)=.3, p$<$.05 & 51  & Chi2  & 12  & NA  & =  & 0.2  & $<$  & 0.05  & 1.0000  & TRUE  & TRUE  & FALSE  \\
   \hline
\end{tabular}
\end{table}

\clearpage

The subsequent test set comprises a vector with 52 example results with all upper and lowercase letters, followed by an appended number two, a number inside brackets, a numerical result and a p-value (e.g.: `A2(12)=.3, p<.05').
The cropped \textit{statcheck} output is displayed in Table \ref{tab::letter2}. 
43 of the 52 input results were detected. The eight example results with the letters D, F, Q, W, Z, r, t, and z, appended with the number 2 are not detected.
Here, all 43 detected results were treated as a chi-square test result, which, in practice, is reasonable for `X2' and `x2'. 
The imprecise detection heuristic for chi-square results treats any plausible report of a model fit (e.g.: $R2(12)=.3, p<.05$) as if it was a report of a chi-square test.

\begin{table}[!ht]
\setlength{\tabcolsep}{4pt}
\caption{Truncated output of: \textit{statcheck(paste0(" ",c(LETTERS,letters)2,"(12)=.3, p<.05"))}}
\label{tab::letter2}
\centering
\scriptsize
\begin{tabular}{rllllllllllllll}
  \hline
 & raw & source & test\_type & df1 & df2 & test\_comp & test\_value & p\_comp & reported\_p & computed\_p & error & decision\_error & one\_tailed\_in\_txt \\
  \hline
1 & A2(12)=.3, p$<$.05 & 01  & Chi2  & 12  & NA  & =  & 0.2  & $<$  & 0.05  & 1  & TRUE  & TRUE  & FALSE  \\
  2 & B2(12)=.3, p$<$.05 & 02  & Chi2  & 12  & NA  & =  & 0.2  & $<$  & 0.05  & 1  & TRUE  & TRUE  & FALSE  \\
  3 & C2(12)=.3, p$<$.05 & 03  & Chi2  & 12  & NA  & =  & 0.2  & $<$  & 0.05  & 1  & TRUE  & TRUE  & FALSE  \\
  4 & E2(12)=.3, p$<$.05 & 05  & Chi2  & 12  & NA  & =  & 0.2  & $<$  & 0.05  & 1  & TRUE  & TRUE  & FALSE  \\
  5 & G2(12)=.3, p$<$.05 & 07  & Chi2  & 12  & NA  & =  & 0.2  & $<$  & 0.05  & 1  & TRUE  & TRUE  & FALSE  \\
...\\
  14 & P2(12)=.3, p$<$.05 & 16  & Chi2  & 12  & NA  & =  & 0.2  & $<$  & 0.05  & 1  & TRUE  & TRUE  & FALSE  \\
  15 & R2(12)=.3, p$<$.05 & 18  & Chi2  & 12  & NA  & =  & 0.2  & $<$  & 0.05  & 1  & TRUE  & TRUE  & FALSE  \\
  16 & S2(12)=.3, p$<$.05 & 19  & Chi2  & 12  & NA  & =  & 0.2  & $<$  & 0.05  & 1  & TRUE  & TRUE  & FALSE  \\
  17 & T2(12)=.3, p$<$.05 & 20  & Chi2  & 12  & NA  & =  & 0.2  & $<$  & 0.05  & 1  & TRUE  & TRUE  & FALSE  \\
  18 & U2(12)=.3, p$<$.05 & 21  & Chi2  & 12  & NA  & =  & 0.2  & $<$  & 0.05  & 1  & TRUE  & TRUE  & FALSE  \\
  19 & V2(12)=.3, p$<$.05 & 22  & Chi2  & 12  & NA  & =  & 0.2  & $<$  & 0.05  & 1  & TRUE  & TRUE  & FALSE  \\
  20 & X2(12)=.3, p$<$.05 & 24  & Chi2  & 12  & NA  & =  & 0.2  & $<$  & 0.05  & 1  & TRUE  & TRUE  & FALSE  \\
  21 & Y2(12)=.3, p$<$.05 & 25  & Chi2  & 12  & NA  & =  & 0.2  & $<$  & 0.05  & 1  & TRUE  & TRUE  & FALSE  \\
  22 & a2(12)=.3, p$<$.05 & 27  & Chi2  & 12  & NA  & =  & 0.2  & $<$  & 0.05  & 1  & TRUE  & TRUE  & FALSE  \\
...\\
  37 & q2(12)=.3, p$<$.05 & 43  & Chi2  & 12  & NA  & =  & 0.2  & $<$  & 0.05  & 1  & TRUE  & TRUE  & FALSE  \\
  38 & s2(12)=.3, p$<$.05 & 45  & Chi2  & 12  & NA  & =  & 0.2  & $<$  & 0.05  & 1  & TRUE  & TRUE  & FALSE  \\
  39 & u2(12)=.3, p$<$.05 & 47  & Chi2  & 12  & NA  & =  & 0.2  & $<$  & 0.05  & 1  & TRUE  & TRUE  & FALSE  \\
  40 & v2(12)=.3, p$<$.05 & 48  & Chi2  & 12  & NA  & =  & 0.2  & $<$  & 0.05  & 1  & TRUE  & TRUE  & FALSE  \\
  41 & w2(12)=.3, p$<$.05 & 49  & Chi2  & 12  & NA  & =  & 0.2  & $<$  & 0.05  & 1  & TRUE  & TRUE  & FALSE  \\
  42 & x2(12)=.3, p$<$.05 & 50  & Chi2  & 12  & NA  & =  & 0.2  & $<$  & 0.05  & 1  & TRUE  & TRUE  & FALSE  \\
  43 & y2(12)=.3, p$<$.05 & 51  & Chi2  & 12  & NA  & =  & 0.2  & $<$  & 0.05  & 1  & TRUE  & TRUE  & FALSE  \\
   \hline
\end{tabular}
\end{table}

To complete the testing procedure with possible simple reports of statistical test results, the next text vector consists of 52 results starting with an upper and lowercase letter followed by the hat sign as an exponent symbol and the number two, a number in brackets, a numerical result and a p-value (e.g.: `A\textasciicircum2(12),.3, p<.05'). 
The cropped \textit{statcheck} output is shown in Table \ref{tab::letterHoch2}. 
Once again, the overly imprecise detection heuristic for chi-square results is apparent. 
It is evident that the handling of the two special characters in the name of a chi-square statistic is not working properly. 
All 52 inputs were detected and handled as chi-square test results, but the raw results extracted by \textit{statcheck} omit the preceding letter of each result. 
Apparently, the implemented search task does not include a letter in front of the exponent symbol. 
This, again, leads to erroneous extractions of plausible reports of model fit tests (e.g.: $R\textasciicircum2(12)=.3, p<.05$) that are mistakenly treated as chi-square test results.

\begin{table}[!ht]
\setlength{\tabcolsep}{4pt}

\caption{Truncated output of: \textit{statcheck(paste0(" ",c(LETTERS,letters)\textasciicircum 2,"(12)=.3, p<.05"))}}
\label{tab::letterHoch2}
\centering
\scriptsize
\begin{tabular}{rllllllllllllll}
  \hline
 & raw & source & test\_type & df1 & df2 & test\_comp & test\_value & p\_comp & reported\_p & computed\_p & error & decision\_error & one\_tailed\_in\_txt\\
  \hline
1 & \verb|^|2(12)=.3, p$<$.05 & 01  & Chi2  & 12  & NA  & =  & 0.2  & $<$  & 0.05  & 1  & TRUE  & TRUE  & FALSE  \\
  2 & \verb|^|2(12)=.3, p$<$.05 & 02  & Chi2  & 12  & NA  & =  & 0.2  & $<$  & 0.05  & 1  & TRUE  & TRUE  & FALSE  \\
...\\
  52 & \verb|^|2(12)=.3, p$<$.05 & 52  & Chi2  & 12  & NA  & =  & 0.2  & $<$  & 0.05  & 1  & TRUE  & TRUE  & FALSE  \\
   \hline
\end{tabular}
\end{table}

\clearpage 

\subsection{Results that \textit{statcheck} does not recognize at all}

Until now, all test results followed a standardized and consistent reporting style consisting of a test statistic with degrees of freedom in brackets (except for the Z-statistic) and a p-value. 
However, in psychology and other disciplines, results are often presented in an individual and non-APA style.

Table \ref{tab::statcheckNoCheck} provides a non-exhaustive inventory of result patterns and one exemplary result, which serves as the last here presented check of \textit{statcheck}. 
None of the listed results is detected. 
Some of these very simple examples supplied here refer to numerous results reported in the literature, as demonstrated earlier \citep{getStats}. 
It is clear that \textit{statcheck} has an overly high false negative rate, which will worsen as researchers increasingly adhere to new reporting guidelines recommending the reporting of effect sizes, confidence intervals, or Bayes factors.

\begin{table}[!ht]
\caption{General result patterns that \textit{statcheck} does not recognize nor spell check and a correspoding simple example}
\centering
\label{tab::statcheckNoCheck}
\begin{tabular}{cp{9cm}l}
\hline
No. & general result pattern & example string\\
\hline
1 & chi-square test results presented with superscripted 2 &  $\chi^2$(12)=3.4, p$<$.05  \\ 
 2 & chi-square with html-coded superscripted 2 &  $\chi<$sup$>$2$<$/sup$>$(12)=3.4, p$<$.05 \\ 
  3 & does not allow a calculation of the p-value &  t=.12, p=.34 \\ 
  4 & p-value only &  p=.12 \\ 
  5 & contains a report of any other value (e.g. Cohen's d) between &  t(12)=1.2, d=3.4, p=.56 \\ 
  6 & is supplied with a semicolon as seperator &  t(12)=1.2; p=.34 \\ 
  7 & is supplied without a seperator &  t(12)=1.2 p=.34 \\ 
  8 & contains the report of the degrees of freedom outside bracket &  t=1.2, df=34, p=.56 \\ 
  9 & contains an exponent or fraction &  t(12)=1.2\verb|^|3, p=4/5 \\ 
  10 & has badly used seperators/decimals/numbers &  t(12)=1..2, p=.n3 \\ 
  11 & has indexed results &  t index(12)=1.2, p=.34 \\ 
  12 & has p-, r- or R\verb|^|2-values outside their valid range &  r(12)=1.2, p=3.45, R\verb|^|2=6.7 \\ 
  13 & has estimatable p-values by beta and standard error &  beta=1.2, SE=.34, p$<$.05 \\ 
  14 & has a value reported with percent sign &  t(12)=1.2, p=5\% \\ 
  15 & a report of multiple results &  all t's(12)$>$1.2, p's$>$.05 \\ 
  16 & is reported with a less equal or greater equal sign or alias &  t(12)$\leq$1.2, p$\geq$.05 \\ 
  17 & has degrees of freedom in squared brackets &  t[12]=1.2, p$<$.05 \\ 
  18 & has high degrees of freedom with coma as punctuation &  t(1,234)=5.6, p$<$.05 \\ 
  19 & has a double space behind the coma &  t(12)=1.2,  p$<$.05 \\ 
  20 & has a space in abbreviation of not significant &  t(12)=.34, n. s.  \\ 
  21 & has badly or non compiled operator due to PDF to text convers &  t(12) 1.2, p 5 .34 \\ 
  \hline
 \end{tabular}
\end{table}

\section{Conclusion}
The ubiquity of R as a statistical tool underscores the need for reliable packages, given its explosive adoption in recent decades \citep{landau2021we}. 
While \textit{statcheck} was developed with the commendable goal of automating the validation of statistical results in scientific articles, its design and implementation significantly restrict its trustworthiness.

It was demonstrated that \textit{statcheck} has serious bugs and does not fulfill scientific requirements as a valid error detector. 
It is neither an appropiate tool for extracting statistical results from text nor a reliable spell checker for results. 
The most critical drawback of \textit{statcheck} is that it neither identifies nor checks most of the results reported in the literature. 
In addition, the implemented extraction heuristic is unsuitable for spell checking reported results, since almost any reporting anomaly results in a non-extraction by \textit{statcheck}. 
The majority of possible typos, reasonable individual reporting styles, and incompletely reported results are not recognized as results at all. 
Unmanageable results are omitted from \textit{statcheck}'s output, so users are more likely to wonder about the non-detection than to get their results checked. 
For authors and journals that strictly follow the manageable APA reporting style, this feature may not be very limiting, but it is highly limiting for the application in most other fields and journals. 
Authors who report effect sizes, which has long been called for, and report them between the test statistic and p-value, will also never get their results checked by \textit{statcheck}.

Looking at the potential input space, \textit{statcheck} exhibits an excessively high rate of both false negatives and false positives. 
In particular, the implemented chi-square detection heuristic suffers from a lack of precision and is too sensitive. 
This may be due to the various possible technical representations of the Greek letter $\chi$ und superscripted 2. 
As a consequence, any squared statistic (e.g. $R^2$) is treated as $\chi^2$ statistic. 
The problem of special character encoding also arises when extracting text from PDF documents and often results in spurious artifacts. 
However, the \textit{statcheck} algorithm is not able to deal with the typical noise that is introduced into text strings extracted from PDF documents.

Undoubtedly, \textit{statcheck} is incapable of detecting a majority of the statistical test results presented in the literature, as it can only process a very restricted range of representations. 
The fact that many statistical results are reported in tabular form (e.g., ANOVA or regression tables) has not been considered here. 
Since it is very uncommon to report tabulated statistical results in a \textit{statcheck}-readable notation, it is obvious that another large number of results actually reported in a manuscript are not detectable by \textit{statcheck}. 

I recommend not using \textit{statcheck}, neither to check the completeness or consistency of statistical test results reported in a single manuscript nor in a body of literature. 
Whereas previously the question was how to deal with the inconsistencies that \textit{statcheck} had correctly identified in the past, now the question is how to deal with the apparent inconsistencies of \textit{statcheck}.

\textit{statcheck}'s limitations cast doubts on the validity of past analyses already carried out with \textit{statcheck}. 
In a recent study, Nuijten et al. \citep{nuijten2023implementing} investigated the effect of using \textit{statcheck} during peer review on a journal-based comparison and found support for a decrease in reporting errors in those journals that had integrated \textit{statcheck} into their review process. 
The results and conclusions of the study sound promising, but the adaptation of the various problems listed here should be considered. 
\textit{statcheck} forces users to report results in a \textit{statcheck}-readable way. 
This may be beneficial in some cases, but it is crucial for plausible other reporting styles. 
It must be considered, that many results pass the required \textit{statcheck} check simply because they remain unrecognized by \textit{statcheck}.

In one-tailed test settings, the test direction must be consistent with the postulated direction. 
It is surely difficult to automatically tell from text whether a specific test result is based on a one-tailed test and whether it is correctly reported in terms of the test direction. 
In standard mode \textit{statcheck}'s option \textit{OneTailedTxt} that searches the text for ``one-sided'', ``one-tailed'', and ``directional'' to identify the possible use of one-sided tests, is deactivated. 
Also its option \textit{OneTailedTests} is deactivated. 
If activated \textit{statcheck} performs its checks on the assumtion that all tests are one-sided.
This is problematic because most of the time not all tests in a manuscript are one-sided and the consistency of the tested direction is never checked.

The more researchers apply directed testing or report corrected p-values, the harder it will be to automatically check results for consistency. 
The challenges in accurately identifying and interpreting one-tailed tests and corrected p-values underscore the complexity of automating statistical checks. 
It remains uncertain whether advancements in AI and machine learning will suffice to address these nuanced tasks effectively.

Despite \textit{statcheck}'s potential utility in specific contexts, its current limitations significantly hinder its broad applicability. 
The responsibility for ensuring the accuracy and plausibility of reported statistical results, therefore, remains predominantly a human endeavor. 
Software tools can aid this process, but they cannot replace the critical eye of a knowledgeable reader  who evaluates the hypotheses tested and conclusions drawn in the context of the study design, adequateness of the analysis and sampling efforts.

Altough it was not my aim to demonstrate the flexibility of the \textit{get.stats} function out of the \textit{JATSdecoder} package, I have processed the full text string with the same 187 arbitrary results and report its output in Table \ref{tab::getstats}. 
All 187 results were detected as sticked results and adequately postprocessed to a matrix containing the categorized numeric standard results, p-values and operators. 
Overall, 184 of the 185 reported p-values were detected correctly. 
The two typos in result 10 in Table \ref{tab::statcheckNoCheck} (`t(12)=1..2, p=.n3') result in a detection of t=1 and a non-detectetion of the p-value. 
For a detailed description of the aim and scope of the \textit{get.stats} function, see \citep{getStats}.

\bibliographystyle{unsrtnat}

\section*{Declarations}
I developed the R package \textit{JATSdecoder} \citep{JATSdecoderCRAN1.2} and the implemented function \textit{get.stats} which enables the extraction of all numerical results in scientific reports in NISO-JATS format without restrictions to reporting style. 

\section*{Contributions}
Ingmar Böschen ran the checks on \textit{statcheck} and wrote the manuscript.

\section*{Conflict of interest} 
The author declares no conflict of interest.

\section*{Availability of code and data}
\begin{itemize}
\item The script to reproduce the check on \textit{statcheck} is stored at: \\
\url{https://github.com/ingmarboeschen/JATSdecoderEvaluation/}
\item Use the there stored docx-file to quickly check \textit{statcheck} at:\\
\url{https://michelenuijten.shinyapps.io/statcheck-web/}

\end{itemize}

\newpage
\section{Appendix}
\begin{table}[ht]
\caption{The output of \textit{JATSdecoder}'s function \textit{get.stats} when applied to the full result string}
\label{tab::getstats}
\begin{adjustbox}{angle=90}
\setlength{\tabcolsep}{2pt}
\centering
\scriptsize
\begin{tabular}{rllrlrlrrlrlrlrlrlrrlrrrrrrlrr}
  \hline
 & result & Z\_op & Z & F\_op & F & t\_op & t & d & r\_op & r & R2\_op & R2 & U\_op & U & H\_op & H & G2\_op & G2 & Chi2 & Q\_op & Q & df1 & df2 & beta & SEbeta & Zest & p\_op & p & recalculatedP \\ 
  \hline
1 & t(12)=2.3, p$<$.05 &  &  &  &  & = & 2.30 &  &  &  &  &  &  &  &  &  &  &  &  &  &  &  & 12.00 &  &  &  & $<$ & 0.05 & 0.04 \\ 
  2 & F(1,23)=4.5, p=.23 &  &  & = & 4.50 &  &  &  &  &  &  &  &  &  &  &  &  &  &  &  &  & 1.00 & 23.00 &  &  &  & = & 0.23 & 0.04 \\ 
  3 & r(12)=.34, p=.56 &  &  &  &  &  &  &  & = & 0.34 &  &  &  &  &  &  &  &  &  &  &  &  & 12.00 &  &  &  & = & 0.56 & 0.23 \\ 
  4 & Z=1.2, p$<$.34 & = & 1.20 &  &  &  &  &  &  &  &  &  &  &  &  &  &  &  &  &  &  &  &  &  &  &  & $<$ & 0.34 & 0.23 \\ 
  5 & Chi\verb|^|2(12)=3.4, p$<$.05 &  &  &  &  &  &  &  &  &  &  &  &  &  &  &  &  &  & 3.40 &  &  & 12.00 &  &  &  &  & $<$ & 0.05 & 0.99 \\ 
  6 & chi2(12)=3.4, p$<$.05 &  &  &  &  &  &  &  &  &  &  &  &  &  &  &  &  &  & 3.40 &  &  & 12.00 &  &  &  &  & $<$ & 0.05 & 0.99 \\ 
  7 & Chi\verb|^|2(12)=3.4, p$<$.05 &  &  &  &  &  &  &  &  &  &  &  &  &  &  &  &  &  & 3.40 &  &  & 12.00 &  &  &  &  & $<$ & 0.05 & 0.99 \\ 
  8 & chi2(12)=3.4, p$<$.05 &  &  &  &  &  &  &  &  &  &  &  &  &  &  &  &  &  & 3.40 &  &  & 12.00 &  &  &  &  & $<$ & 0.05 & 0.99 \\ 
  9 & Q(12)=3.4, p$<$.01 &  &  &  &  &  &  &  &  &  &  &  &  &  &  &  &  &  &  & = & 3.40 & 12.00 &  &  &  &  & $<$ & 0.01 & 0.99 \\ 
  10 & t(12)=.34 &  &  &  &  & = & 0.34 &  &  &  &  &  &  &  &  &  &  &  &  &  &  &  & 12.00 &  &  &  &  &  & 0.74 \\ 
  11 & A(12)=.3, p$<$.05 &  &  &  &  &  &  &  &  &  &  &  &  &  &  &  &  &  &  &  &  &  &  &  &  &  & $<$ & 0.05 &  \\ 
  12 & B(12)=.3, p$<$.05 &  &  &  &  &  &  &  &  &  &  &  &  &  &  &  &  &  &  &  &  &  &  &  &  &  & $<$ & 0.05 &  \\ 
  13 & C(12)=.3, p$<$.05 &  &  &  &  &  &  &  &  &  &  &  &  &  &  &  &  &  &  &  &  &  &  &  &  &  & $<$ & 0.05 &  \\ 
  14 & D(12)=.3, p$<$.05 &  &  &  &  &  &  &  &  &  &  &  &  &  &  &  &  &  &  &  &  &  &  &  &  &  & $<$ & 0.05 &  \\ 
  15 & E(12)=.3, p$<$.05 &  &  &  &  &  &  &  &  &  &  &  &  &  &  &  &  &  &  &  &  &  &  &  &  &  & $<$ & 0.05 &  \\ 
  16 & F(12)=.3, p$<$.05 &  &  &  &  &  &  &  &  &  &  &  &  &  &  &  &  &  &  &  &  &  &  &  &  &  & $<$ & 0.05 &  \\ 
  17 & G(12)=.3, p$<$.05 &  &  &  &  &  &  &  &  &  &  &  &  &  &  &  &  &  &  &  &  &  &  &  &  &  & $<$ & 0.05 &  \\ 
  18 & H(12)=.3, p$<$.05 &  &  &  &  &  &  &  &  &  &  &  &  &  & = & 0.30 &  &  &  &  &  & 12.00 &  &  &  &  & $<$ & 0.05 & 1.00 \\ 
  19 & I(12)=.3, p$<$.05 &  &  &  &  &  &  &  &  &  &  &  &  &  &  &  &  &  &  &  &  &  &  &  &  &  & $<$ & 0.05 &  \\ 
  20 & J(12)=.3, p$<$.05 &  &  &  &  &  &  &  &  &  &  &  &  &  &  &  &  &  &  &  &  &  &  &  &  &  & $<$ & 0.05 &  \\ 
  21 & K(12)=.3, p$<$.05 &  &  &  &  &  &  &  &  &  &  &  &  &  &  &  &  &  &  &  &  &  &  &  &  &  & $<$ & 0.05 &  \\ 
  22 & L(12)=.3, p$<$.05 &  &  &  &  &  &  &  &  &  &  &  &  &  &  &  &  &  &  &  &  &  &  &  &  &  & $<$ & 0.05 &  \\ 
  23 & M(12)=.3, p$<$.05 &  &  &  &  &  &  &  &  &  &  &  &  &  &  &  &  &  &  &  &  &  &  &  &  &  & $<$ & 0.05 &  \\ 
  24 & N(12)=.3, p$<$.05 &  &  &  &  &  &  &  &  &  &  &  &  &  &  &  &  &  &  &  &  &  &  &  &  &  & $<$ & 0.05 &  \\ 
  25 & O(12)=.3, p$<$.05 &  &  &  &  &  &  &  &  &  &  &  &  &  &  &  &  &  &  &  &  &  &  &  &  &  & $<$ & 0.05 &  \\ 
  26 & P(12)=.3, p$<$.05 &  &  &  &  &  &  &  &  &  &  &  &  &  &  &  &  &  &  &  &  &  &  &  &  &  & $<$ & 0.05 &  \\ 
  27 & Q(12)=.3, p$<$.05 &  &  &  &  &  &  &  &  &  &  &  &  &  &  &  &  &  &  & = & 0.30 & 12.00 &  &  &  &  & $<$ & 0.05 & 1.00 \\ 
  28 & R(12)=.3, p$<$.05 &  &  &  &  &  &  &  &  &  &  &  &  &  &  &  &  &  &  &  &  &  &  &  &  &  & $<$ & 0.05 &  \\ 
  29 & S(12)=.3, p$<$.05 &  &  &  &  &  &  &  &  &  &  &  &  &  &  &  &  &  &  &  &  &  &  &  &  &  & $<$ & 0.05 &  \\ 
  30 & T(12)=.3, p$<$.05 &  &  &  &  &  &  &  &  &  &  &  &  &  &  &  &  &  &  &  &  &  &  &  &  &  & $<$ & 0.05 &  \\ 
  31 & U(12)=.3, p$<$.05 &  &  &  &  &  &  &  &  &  &  &  & = & 0.30 &  &  &  &  &  &  &  &  &  &  &  &  & $<$ & 0.05 &  \\ 
  32 & V(12)=.3, p$<$.05 &  &  &  &  &  &  &  &  &  &  &  &  &  &  &  &  &  &  &  &  &  &  &  &  &  & $<$ & 0.05 &  \\ 
  33 & W(12)=.3, p$<$.05 &  &  &  &  &  &  &  &  &  &  &  &  &  &  &  &  &  &  &  &  &  &  &  &  &  & $<$ & 0.05 &  \\ 
  34 & X(12)=.3, p$<$.05 &  &  &  &  &  &  &  &  &  &  &  &  &  &  &  &  &  & 0.30 &  &  & 12.00 &  &  &  &  & $<$ & 0.05 & 1.00 \\ 
  35 & Y(12)=.3, p$<$.05 &  &  &  &  &  &  &  &  &  &  &  &  &  &  &  &  &  &  &  &  &  &  &  &  &  & $<$ & 0.05 &  \\ 
  36 & Z(12)=.3, p$<$.05 & = & 0.30 &  &  &  &  &  &  &  &  &  &  &  &  &  &  &  &  &  &  &  &  &  &  &  & $<$ & 0.05 & 0.76 \\ 
  37 & a(12)=.3, p$<$.05 &  &  &  &  &  &  &  &  &  &  &  &  &  &  &  &  &  &  &  &  &  &  &  &  &  & $<$ & 0.05 &  \\ 
  38 & b(12)=.3, p$<$.05 &  &  &  &  &  &  &  &  &  &  &  &  &  &  &  &  &  &  &  &  &  &  &  &  &  & $<$ & 0.05 &  \\ 
  39 & c(12)=.3, p$<$.05 &  &  &  &  &  &  &  &  &  &  &  &  &  &  &  &  &  &  &  &  &  &  &  &  &  & $<$ & 0.05 &  \\ 
  40 & d(12)=.3, p$<$.05 &  &  &  &  &  &  &  &  &  &  &  &  &  &  &  &  &  &  &  &  &  &  &  &  &  & $<$ & 0.05 &  \\ 
  41 & e(12)=.3, p$<$.05 &  &  &  &  &  &  &  &  &  &  &  &  &  &  &  &  &  &  &  &  &  &  &  &  &  & $<$ & 0.05 &  \\ 
  42 & f(12)=.3, p$<$.05 &  &  &  &  &  &  &  &  &  &  &  &  &  &  &  &  &  &  &  &  &  &  &  &  &  & $<$ & 0.05 &  \\ 
  43 & g(12)=.3, p$<$.05 &  &  &  &  &  &  &  &  &  &  &  &  &  &  &  &  &  &  &  &  &  &  &  &  &  & $<$ & 0.05 &  \\ 
  44 & h(12)=.3, p$<$.05 &  &  &  &  &  &  &  &  &  &  &  &  &  &  &  &  &  &  &  &  &  &  &  &  &  & $<$ & 0.05 &  \\ 
  45 & i(12)=.3, p$<$.05 &  &  &  &  &  &  &  &  &  &  &  &  &  &  &  &  &  &  &  &  &  &  &  &  &  & $<$ & 0.05 &  \\ 
  46 & j(12)=.3, p$<$.05 &  &  &  &  &  &  &  &  &  &  &  &  &  &  &  &  &  &  &  &  &  &  &  &  &  & $<$ & 0.05 &  \\ 
  47 & k(12)=.3, p$<$.05 &  &  &  &  &  &  &  &  &  &  &  &  &  &  &  &  &  &  &  &  &  &  &  &  &  & $<$ & 0.05 &  \\ 
  48 & l(12)=.3, p$<$.05 &  &  &  &  &  &  &  &  &  &  &  &  &  &  &  &  &  &  &  &  &  &  &  &  &  & $<$ & 0.05 &  \\ 
  49 & m(12)=.3, p$<$.05 &  &  &  &  &  &  &  &  &  &  &  &  &  &  &  &  &  &  &  &  &  &  &  &  &  & $<$ & 0.05 &  \\ 
  50 & n(12)=.3, p$<$.05 &  &  &  &  &  &  &  &  &  &  &  &  &  &  &  &  &  &  &  &  &  &  &  &  &  & $<$ & 0.05 &  \\ 
  51 & o(12)=.3, p$<$.05 &  &  &  &  &  &  &  &  &  &  &  &  &  &  &  &  &  &  &  &  &  &  &  &  &  & $<$ & 0.05 &  \\ 
  52 & p(12)=.3, p$<$.05 &  &  &  &  &  &  &  &  &  &  &  &  &  &  &  &  &  &  &  &  &  &  &  &  &  & $<$ & 0.05 &  \\ 
  53 & q(12)=.3, p$<$.05 &  &  &  &  &  &  &  &  &  &  &  &  &  &  &  &  &  &  & = & 0.30 & 12.00 &  &  &  &  & $<$ & 0.05 & 1.00 \\ 
  54 & r(12)=.3, p$<$.05 &  &  &  &  &  &  &  & = & 0.30 &  &  &  &  &  &  &  &  &  &  &  &  & 12.00 &  &  &  & $<$ & 0.05 & 0.30 \\ 
  55 & s(12)=.3, p$<$.05 &  &  &  &  &  &  &  &  &  &  &  &  &  &  &  &  &  &  &  &  &  &  &  &  &  & $<$ & 0.05 &  \\ 
  56 & t(12)=.3, p$<$.05 &  &  &  &  & = & 0.30 &  &  &  &  &  &  &  &  &  &  &  &  &  &  &  & 12.00 &  &  &  & $<$ & 0.05 & 0.77 \\ 
  57 & u(12)=.3, p$<$.05 &  &  &  &  &  &  &  &  &  &  &  &  &  &  &  &  &  &  &  &  &  &  &  &  &  & $<$ & 0.05 &  \\ 
  58 & v(12)=.3, p$<$.05 &  &  &  &  &  &  &  &  &  &  &  &  &  &  &  &  &  &  &  &  &  &  &  &  &  & $<$ & 0.05 &  \\ 
  59 & w(12)=.3, p$<$.05 &  &  &  &  &  &  &  &  &  &  &  &  &  &  &  &  &  &  &  &  &  &  &  &  &  & $<$ & 0.05 &  \\ 
  60 & x(12)=.3, p$<$.05 &  &  &  &  &  &  &  &  &  &  &  &  &  &  &  &  &  & 0.30 &  &  & 12.00 &  &  &  &  & $<$ & 0.05 & 1.00 \\ 
  61 & y(12)=.3, p$<$.05 &  &  &  &  &  &  &  &  &  &  &  &  &  &  &  &  &  &  &  &  &  &  &  &  &  & $<$ & 0.05 &  \\ 
   \hline
\end{tabular}
\end{adjustbox}
\end{table}

\begin{table}[ht]
  \begin{adjustbox}{angle=90}
\setlength{\tabcolsep}{2pt}
\centering
\scriptsize
\begin{tabular}{rllrlrlrrlrlrlrlrlrrlrrrrrrlrr}
  \hline
 & result & Z\_op & Z & F\_op & F & t\_op & t & d & r\_op & r & R2\_op & R2 & U\_op & U & H\_op & H & G2\_op & G2 & Chi2 & Q\_op & Q & df1 & df2 & beta & SEbeta & Zest & p\_op & p & recalculatedP \\ 
  \hline

  62 & z(12)=.3, p$<$.05 & = & 0.30 &  &  &  &  &  &  &  &  &  &  &  &  &  &  &  &  &  &  &  &  &  &  &  & $<$ & 0.05 & 0.76 \\ 
  63 & A2(12)=.3, p$<$.05 &  &  &  &  &  &  &  &  &  &  &  &  &  &  &  &  &  &  &  &  &  &  &  &  &  & $<$ & 0.05 &  \\ 
  64 & B2(12)=.3, p$<$.05 &  &  &  &  &  &  &  &  &  &  &  &  &  &  &  &  &  &  &  &  &  &  &  &  &  & $<$ & 0.05 &  \\ 
  65 & C2(12)=.3, p$<$.05 &  &  &  &  &  &  &  &  &  &  &  &  &  &  &  &  &  &  &  &  &  &  &  &  &  & $<$ & 0.05 &  \\ 
  66 & D2(12)=.3, p$<$.05 &  &  &  &  &  &  &  &  &  &  &  &  &  &  &  &  &  &  &  &  &  &  &  &  &  & $<$ & 0.05 &  \\ 
  67 & E2(12)=.3, p$<$.05 &  &  &  &  &  &  &  &  &  &  &  &  &  &  &  &  &  &  &  &  &  &  &  &  &  & $<$ & 0.05 &  \\ 
  68 & F2(12)=.3, p$<$.05 &  &  &  &  &  &  &  &  &  &  &  &  &  &  &  &  &  &  &  &  &  &  &  &  &  & $<$ & 0.05 &  \\ 
  69 & G2(12)=.3, p$<$.05 &  &  &  &  &  &  &  &  &  &  &  &  &  &  &  & = & 0.30 &  &  &  & 12.00 &  &  &  &  & $<$ & 0.05 & 1.00 \\ 
  70 & H2(12)=.3, p$<$.05 &  &  &  &  &  &  &  &  &  &  &  &  &  & = & 0.30 &  &  &  &  &  & 12.00 &  &  &  &  & $<$ & 0.05 & 1.00 \\ 
  71 & I2(12)=.3, p$<$.05 &  &  &  &  &  &  &  &  &  &  &  &  &  &  &  &  &  &  &  &  &  &  &  &  &  & $<$ & 0.05 &  \\ 
  72 & J2(12)=.3, p$<$.05 &  &  &  &  &  &  &  &  &  &  &  &  &  &  &  &  &  &  &  &  &  &  &  &  &  & $<$ & 0.05 &  \\ 
  73 & K2(12)=.3, p$<$.05 &  &  &  &  &  &  &  &  &  &  &  &  &  &  &  &  &  &  &  &  &  &  &  &  &  & $<$ & 0.05 &  \\ 
  74 & L2(12)=.3, p$<$.05 &  &  &  &  &  &  &  &  &  &  &  &  &  &  &  &  &  &  &  &  &  &  &  &  &  & $<$ & 0.05 &  \\ 
  75 & M2(12)=.3, p$<$.05 &  &  &  &  &  &  &  &  &  &  &  &  &  &  &  &  &  &  &  &  &  &  &  &  &  & $<$ & 0.05 &  \\ 
  76 & N2(12)=.3, p$<$.05 &  &  &  &  &  &  &  &  &  &  &  &  &  &  &  &  &  &  &  &  &  &  &  &  &  & $<$ & 0.05 &  \\ 
  77 & O2(12)=.3, p$<$.05 &  &  &  &  &  &  &  &  &  &  &  &  &  &  &  &  &  &  &  &  &  &  &  &  &  & $<$ & 0.05 &  \\ 
  78 & P2(12)=.3, p$<$.05 &  &  &  &  &  &  &  &  &  &  &  &  &  &  &  &  &  &  &  &  &  &  &  &  &  & $<$ & 0.05 &  \\ 
  79 & Q2(12)=.3, p$<$.05 &  &  &  &  &  &  &  &  &  &  &  &  &  &  &  &  &  &  & = & 0.30 & 12.00 &  &  &  &  & $<$ & 0.05 & 1.00 \\ 
  80 & R2(12)=.3, p$<$.05 &  &  &  &  &  &  &  &  &  & = & 0.30 &  &  &  &  &  &  &  &  &  &  &  &  &  &  & $<$ & 0.05 &  \\ 
  81 & S2(12)=.3, p$<$.05 &  &  &  &  &  &  &  &  &  &  &  &  &  &  &  &  &  &  &  &  &  &  &  &  &  & $<$ & 0.05 &  \\ 
  82 & T2(12)=.3, p$<$.05 &  &  &  &  &  &  &  &  &  &  &  &  &  &  &  &  &  &  &  &  &  &  &  &  &  & $<$ & 0.05 &  \\ 
  83 & U2(12)=.3, p$<$.05 &  &  &  &  &  &  &  &  &  &  &  &  &  &  &  &  &  &  &  &  &  &  &  &  &  & $<$ & 0.05 &  \\ 
  84 & V2(12)=.3, p$<$.05 &  &  &  &  &  &  &  &  &  &  &  &  &  &  &  &  &  &  &  &  &  &  &  &  &  & $<$ & 0.05 &  \\ 
  85 & W2(12)=.3, p$<$.05 &  &  &  &  &  &  &  &  &  &  &  &  &  &  &  &  &  &  &  &  &  &  &  &  &  & $<$ & 0.05 &  \\ 
  86 & chi2(12)=.3, p$<$.05 &  &  &  &  &  &  &  &  &  &  &  &  &  &  &  &  &  & 0.30 &  &  & 12.00 &  &  &  &  & $<$ & 0.05 & 1.00 \\ 
  87 & Y2(12)=.3, p$<$.05 &  &  &  &  &  &  &  &  &  &  &  &  &  &  &  &  &  &  &  &  &  &  &  &  &  & $<$ & 0.05 &  \\ 
  88 & Z2(12)=.3, p$<$.05 &  &  &  &  &  &  &  &  &  &  &  &  &  &  &  &  &  &  &  &  &  &  &  &  &  & $<$ & 0.05 &  \\ 
  89 & a2(12)=.3, p$<$.05 &  &  &  &  &  &  &  &  &  &  &  &  &  &  &  &  &  &  &  &  &  &  &  &  &  & $<$ & 0.05 &  \\ 
  90 & b2(12)=.3, p$<$.05 &  &  &  &  &  &  &  &  &  &  &  &  &  &  &  &  &  &  &  &  &  &  &  &  &  & $<$ & 0.05 &  \\ 
  91 & c2(12)=.3, p$<$.05 &  &  &  &  &  &  &  &  &  &  &  &  &  &  &  &  &  &  &  &  &  &  &  &  &  & $<$ & 0.05 &  \\ 
  92 & d2(12)=.3, p$<$.05 &  &  &  &  &  &  &  &  &  &  &  &  &  &  &  &  &  &  &  &  &  &  &  &  &  & $<$ & 0.05 &  \\ 
  93 & e2(12)=.3, p$<$.05 &  &  &  &  &  &  &  &  &  &  &  &  &  &  &  &  &  &  &  &  &  &  &  &  &  & $<$ & 0.05 &  \\ 
  94 & f2(12)=.3, p$<$.05 &  &  &  &  &  &  &  &  &  &  &  &  &  &  &  &  &  &  &  &  &  &  &  &  &  & $<$ & 0.05 &  \\ 
  95 & g2(12)=.3, p$<$.05 &  &  &  &  &  &  &  &  &  &  &  &  &  &  &  &  &  &  &  &  &  &  &  &  &  & $<$ & 0.05 &  \\ 
  96 & h2(12)=.3, p$<$.05 &  &  &  &  &  &  &  &  &  &  &  &  &  &  &  &  &  &  &  &  &  &  &  &  &  & $<$ & 0.05 &  \\ 
  97 & i2(12)=.3, p$<$.05 &  &  &  &  &  &  &  &  &  &  &  &  &  &  &  &  &  &  &  &  &  &  &  &  &  & $<$ & 0.05 &  \\ 
  98 & j2(12)=.3, p$<$.05 &  &  &  &  &  &  &  &  &  &  &  &  &  &  &  &  &  &  &  &  &  &  &  &  &  & $<$ & 0.05 &  \\ 
  99 & k2(12)=.3, p$<$.05 &  &  &  &  &  &  &  &  &  &  &  &  &  &  &  &  &  &  &  &  &  &  &  &  &  & $<$ & 0.05 &  \\ 
  100 & l2(12)=.3, p$<$.05 &  &  &  &  &  &  &  &  &  &  &  &  &  &  &  &  &  &  &  &  &  &  &  &  &  & $<$ & 0.05 &  \\ 
  101 & m2(12)=.3, p$<$.05 &  &  &  &  &  &  &  &  &  &  &  &  &  &  &  &  &  &  &  &  &  &  &  &  &  & $<$ & 0.05 &  \\ 
  102 & n2(12)=.3, p$<$.05 &  &  &  &  &  &  &  &  &  &  &  &  &  &  &  &  &  &  &  &  &  &  &  &  &  & $<$ & 0.05 &  \\ 
  103 & o2(12)=.3, p$<$.05 &  &  &  &  &  &  &  &  &  &  &  &  &  &  &  &  &  &  &  &  &  &  &  &  &  & $<$ & 0.05 &  \\ 
  104 & p2(12)=.3, p$<$.05 &  &  &  &  &  &  &  &  &  &  &  &  &  &  &  &  &  &  &  &  &  &  &  &  &  & $<$ & 0.05 &  \\ 
  105 & q2(12)=.3, p$<$.05 &  &  &  &  &  &  &  &  &  &  &  &  &  &  &  &  &  &  &  &  &  &  &  &  &  & $<$ & 0.05 &  \\ 
  106 & r2(12)=.3, p$<$.05 &  &  &  &  &  &  &  &  &  & = & 0.30 &  &  &  &  &  &  &  &  &  &  &  &  &  &  & $<$ & 0.05 &  \\ 
  107 & s2(12)=.3, p$<$.05 &  &  &  &  &  &  &  &  &  &  &  &  &  &  &  &  &  &  &  &  &  &  &  &  &  & $<$ & 0.05 &  \\ 
  108 & t2(12)=.3, p$<$.05 &  &  &  &  & = & 0.30 &  &  &  &  &  &  &  &  &  &  &  &  &  &  &  & 12.00 &  &  &  & $<$ & 0.05 & 0.77 \\ 
  109 & u2(12)=.3, p$<$.05 &  &  &  &  &  &  &  &  &  &  &  &  &  &  &  &  &  &  &  &  &  &  &  &  &  & $<$ & 0.05 &  \\ 
  110 & v2(12)=.3, p$<$.05 &  &  &  &  &  &  &  &  &  &  &  &  &  &  &  &  &  &  &  &  &  &  &  &  &  & $<$ & 0.05 &  \\ 
  111 & w2(12)=.3, p$<$.05 &  &  &  &  &  &  &  &  &  &  &  &  &  &  &  &  &  &  &  &  &  &  &  &  &  & $<$ & 0.05 &  \\ 
  112 & chi2(12)=.3, p$<$.05 &  &  &  &  &  &  &  &  &  &  &  &  &  &  &  &  &  & 0.30 &  &  & 12.00 &  &  &  &  & $<$ & 0.05 & 1.00 \\ 
  113 & y2(12)=.3, p$<$.05 &  &  &  &  &  &  &  &  &  &  &  &  &  &  &  &  &  &  &  &  &  &  &  &  &  & $<$ & 0.05 &  \\ 
  114 & z2(12)=.3, p$<$.05 &  &  &  &  &  &  &  &  &  &  &  &  &  &  &  &  &  &  &  &  &  &  &  &  &  & $<$ & 0.05 &  \\ 
  115 & A\verb|^|2(12)=.3, p$<$.05 &  &  &  &  &  &  &  &  &  &  &  &  &  &  &  &  &  &  &  &  &  &  &  &  &  & $<$ & 0.05 &  \\ 
  116 & B\verb|^|2(12)=.3, p$<$.05 &  &  &  &  &  &  &  &  &  &  &  &  &  &  &  &  &  &  &  &  &  &  &  &  &  & $<$ & 0.05 &  \\ 
  117 & C\verb|^|2(12)=.3, p$<$.05 &  &  &  &  &  &  &  &  &  &  &  &  &  &  &  &  &  &  &  &  &  &  &  &  &  & $<$ & 0.05 &  \\ 
  118 & D\verb|^|2(12)=.3, p$<$.05 &  &  &  &  &  &  &  &  &  &  &  &  &  &  &  &  &  &  &  &  &  &  &  &  &  & $<$ & 0.05 &  \\ 
  119 & E\verb|^|2(12)=.3, p$<$.05 &  &  &  &  &  &  &  &  &  &  &  &  &  &  &  &  &  &  &  &  &  &  &  &  &  & $<$ & 0.05 &  \\ 
  120 & F\verb|^|2(12)=.3, p$<$.05 &  &  &  &  &  &  &  &  &  &  &  &  &  &  &  &  &  &  &  &  &  &  &  &  &  & $<$ & 0.05 &  \\ 
  121 & G\verb|^|2(12)=.3, p$<$.05 &  &  &  &  &  &  &  &  &  &  &  &  &  &  &  & = & 0.30 &  &  &  & 12.00 &  &  &  &  & $<$ & 0.05 & 1.00 \\ 
  122 & H\verb|^|2(12)=.3, p$<$.05 &  &  &  &  &  &  &  &  &  &  &  &  &  & = & 0.30 &  &  &  &  &  & 12.00 &  &  &  &  & $<$ & 0.05 & 1.00 \\ 
  123 & I\verb|^|2(12)=.3, p$<$.05 &  &  &  &  &  &  &  &  &  &  &  &  &  &  &  &  &  &  &  &  &  &  &  &  &  & $<$ & 0.05 &  \\ 
   \hline
\end{tabular}
\end{adjustbox}
\end{table}

\begin{table}[ht]
\begin{adjustbox}{angle=90}
\setlength{\tabcolsep}{2pt}
\centering
\scriptsize
\begin{tabular}{rllrlrlrrlrlrlrlrlrrlrrrrrrlrr}
  \hline
 & result & Z\_op & Z & F\_op & F & t\_op & t & d & r\_op & r & R2\_op & R2 & U\_op & U & H\_op & H & G2\_op & G2 & Chi2 & Q\_op & Q & df1 & df2 & beta & SEbeta & Zest & p\_op & p & recalculatedP \\ 
  \hline
  124 & J\verb|^|2(12)=.3, p$<$.05 &  &  &  &  &  &  &  &  &  &  &  &  &  &  &  &  &  &  &  &  &  &  &  &  &  & $<$ & 0.05 &  \\ 
  125 & K\verb|^|2(12)=.3, p$<$.05 &  &  &  &  &  &  &  &  &  &  &  &  &  &  &  &  &  &  &  &  &  &  &  &  &  & $<$ & 0.05 &  \\ 
  126 & L\verb|^|2(12)=.3, p$<$.05 &  &  &  &  &  &  &  &  &  &  &  &  &  &  &  &  &  &  &  &  &  &  &  &  &  & $<$ & 0.05 &  \\ 
  127 & M\verb|^|2(12)=.3, p$<$.05 &  &  &  &  &  &  &  &  &  &  &  &  &  &  &  &  &  &  &  &  &  &  &  &  &  & $<$ & 0.05 &  \\ 
  128 & N\verb|^|2(12)=.3, p$<$.05 &  &  &  &  &  &  &  &  &  &  &  &  &  &  &  &  &  &  &  &  &  &  &  &  &  & $<$ & 0.05 &  \\ 
  129 & O\verb|^|2(12)=.3, p$<$.05 &  &  &  &  &  &  &  &  &  &  &  &  &  &  &  &  &  &  &  &  &  &  &  &  &  & $<$ & 0.05 &  \\ 
  130 & P\verb|^|2(12)=.3, p$<$.05 &  &  &  &  &  &  &  &  &  &  &  &  &  &  &  &  &  &  &  &  &  &  &  &  &  & $<$ & 0.05 &  \\ 
  131 & Q\verb|^|2(12)=.3, p$<$.05 &  &  &  &  &  &  &  &  &  &  &  &  &  &  &  &  &  &  & = & 0.30 & 12.00 &  &  &  &  & $<$ & 0.05 & 1.00 \\ 
  132 & R\verb|^|2(12)=.3, p$<$.05 &  &  &  &  &  &  &  &  &  & = & 0.30 &  &  &  &  &  &  &  &  &  &  &  &  &  &  & $<$ & 0.05 &  \\ 
  133 & S\verb|^|2(12)=.3, p$<$.05 &  &  &  &  &  &  &  &  &  &  &  &  &  &  &  &  &  &  &  &  &  &  &  &  &  & $<$ & 0.05 &  \\ 
  134 & T\verb|^|2(12)=.3, p$<$.05 &  &  &  &  &  &  &  &  &  &  &  &  &  &  &  &  &  &  &  &  &  &  &  &  &  & $<$ & 0.05 &  \\ 
  135 & U\verb|^|2(12)=.3, p$<$.05 &  &  &  &  &  &  &  &  &  &  &  &  &  &  &  &  &  &  &  &  &  &  &  &  &  & $<$ & 0.05 &  \\ 
  136 & V\verb|^|2(12)=.3, p$<$.05 &  &  &  &  &  &  &  &  &  &  &  &  &  &  &  &  &  &  &  &  &  &  &  &  &  & $<$ & 0.05 &  \\ 
  137 & W\verb|^|2(12)=.3, p$<$.05 &  &  &  &  &  &  &  &  &  &  &  &  &  &  &  &  &  &  &  &  &  &  &  &  &  & $<$ & 0.05 &  \\ 
  138 & chi2(12)=.3, p$<$.05 &  &  &  &  &  &  &  &  &  &  &  &  &  &  &  &  &  & 0.30 &  &  & 12.00 &  &  &  &  & $<$ & 0.05 & 1.00 \\ 
  139 & Y\verb|^|2(12)=.3, p$<$.05 &  &  &  &  &  &  &  &  &  &  &  &  &  &  &  &  &  &  &  &  &  &  &  &  &  & $<$ & 0.05 &  \\ 
  140 & Z\verb|^|2(12)=.3, p$<$.05 &  &  &  &  &  &  &  &  &  &  &  &  &  &  &  &  &  &  &  &  &  &  &  &  &  & $<$ & 0.05 &  \\ 
  141 & a\verb|^|2(12)=.3, p$<$.05 &  &  &  &  &  &  &  &  &  &  &  &  &  &  &  &  &  &  &  &  &  &  &  &  &  & $<$ & 0.05 &  \\ 
  142 & b\verb|^|2(12)=.3, p$<$.05 &  &  &  &  &  &  &  &  &  &  &  &  &  &  &  &  &  &  &  &  &  &  &  &  &  & $<$ & 0.05 &  \\ 
  143 & c\verb|^|2(12)=.3, p$<$.05 &  &  &  &  &  &  &  &  &  &  &  &  &  &  &  &  &  &  &  &  &  &  &  &  &  & $<$ & 0.05 &  \\ 
  144 & d\verb|^|2(12)=.3, p$<$.05 &  &  &  &  &  &  &  &  &  &  &  &  &  &  &  &  &  &  &  &  &  &  &  &  &  & $<$ & 0.05 &  \\ 
  145 & e\verb|^|2(12)=.3, p$<$.05 &  &  &  &  &  &  &  &  &  &  &  &  &  &  &  &  &  &  &  &  &  &  &  &  &  & $<$ & 0.05 &  \\ 
  146 & f\verb|^|2(12)=.3, p$<$.05 &  &  &  &  &  &  &  &  &  &  &  &  &  &  &  &  &  &  &  &  &  &  &  &  &  & $<$ & 0.05 &  \\ 
  147 & g\verb|^|2(12)=.3, p$<$.05 &  &  &  &  &  &  &  &  &  &  &  &  &  &  &  &  &  &  &  &  &  &  &  &  &  & $<$ & 0.05 &  \\ 
  148 & h\verb|^|2(12)=.3, p$<$.05 &  &  &  &  &  &  &  &  &  &  &  &  &  &  &  &  &  &  &  &  &  &  &  &  &  & $<$ & 0.05 &  \\ 
  149 & i\verb|^|2(12)=.3, p$<$.05 &  &  &  &  &  &  &  &  &  &  &  &  &  &  &  &  &  &  &  &  &  &  &  &  &  & $<$ & 0.05 &  \\ 
  150 & j\verb|^|2(12)=.3, p$<$.05 &  &  &  &  &  &  &  &  &  &  &  &  &  &  &  &  &  &  &  &  &  &  &  &  &  & $<$ & 0.05 &  \\ 
  151 & k\verb|^|2(12)=.3, p$<$.05 &  &  &  &  &  &  &  &  &  &  &  &  &  &  &  &  &  &  &  &  &  &  &  &  &  & $<$ & 0.05 &  \\ 
  152 & l\verb|^|2(12)=.3, p$<$.05 &  &  &  &  &  &  &  &  &  &  &  &  &  &  &  &  &  &  &  &  &  &  &  &  &  & $<$ & 0.05 &  \\ 
  153 & m\verb|^|2(12)=.3, p$<$.05 &  &  &  &  &  &  &  &  &  &  &  &  &  &  &  &  &  &  &  &  &  &  &  &  &  & $<$ & 0.05 &  \\ 
  154 & n\verb|^|2(12)=.3, p$<$.05 &  &  &  &  &  &  &  &  &  &  &  &  &  &  &  &  &  &  &  &  &  &  &  &  &  & $<$ & 0.05 &  \\ 
  155 & o\verb|^|2(12)=.3, p$<$.05 &  &  &  &  &  &  &  &  &  &  &  &  &  &  &  &  &  &  &  &  &  &  &  &  &  & $<$ & 0.05 &  \\ 
  156 & p\verb|^|2(12)=.3, p$<$.05 &  &  &  &  &  &  &  &  &  &  &  &  &  &  &  &  &  &  &  &  &  &  &  &  &  & $<$ & 0.05 &  \\ 
  157 & q\verb|^|2(12)=.3, p$<$.05 &  &  &  &  &  &  &  &  &  &  &  &  &  &  &  &  &  &  &  &  &  &  &  &  &  & $<$ & 0.05 &  \\ 
  158 & r\verb|^|2(12)=.3, p$<$.05 &  &  &  &  &  &  &  &  &  & = & 0.30 &  &  &  &  &  &  &  &  &  &  &  &  &  &  & $<$ & 0.05 &  \\ 
  159 & s\verb|^|2(12)=.3, p$<$.05 &  &  &  &  &  &  &  &  &  &  &  &  &  &  &  &  &  &  &  &  &  &  &  &  &  & $<$ & 0.05 &  \\ 
  160 & t\verb|^|2(12)=.3, p$<$.05 &  &  &  &  & = & 0.30 &  &  &  &  &  &  &  &  &  &  &  &  &  &  &  & 12.00 &  &  &  & $<$ & 0.05 & 0.77 \\ 
  161 & u\verb|^|2(12)=.3, p$<$.05 &  &  &  &  &  &  &  &  &  &  &  &  &  &  &  &  &  &  &  &  &  &  &  &  &  & $<$ & 0.05 &  \\ 
  162 & v\verb|^|2(12)=.3, p$<$.05 &  &  &  &  &  &  &  &  &  &  &  &  &  &  &  &  &  &  &  &  &  &  &  &  &  & $<$ & 0.05 &  \\ 
  163 & w\verb|^|2(12)=.3, p$<$.05 &  &  &  &  &  &  &  &  &  &  &  &  &  &  &  &  &  &  &  &  &  &  &  &  &  & $<$ & 0.05 &  \\ 
  164 & chi2(12)=.3, p$<$.05 &  &  &  &  &  &  &  &  &  &  &  &  &  &  &  &  &  & 0.30 &  &  & 12.00 &  &  &  &  & $<$ & 0.05 & 1.00 \\ 
  165 & y\verb|^|2(12)=.3, p$<$.05 &  &  &  &  &  &  &  &  &  &  &  &  &  &  &  &  &  &  &  &  &  &  &  &  &  & $<$ & 0.05 &  \\ 
  166 & z\verb|^|2(12)=.3, p$<$.05 &  &  &  &  &  &  &  &  &  &  &  &  &  &  &  &  &  &  &  &  &  &  &  &  &  & $<$ & 0.05 &  \\ 
  167 & chi2(12)=3.4, p$<$.05 &  &  &  &  &  &  &  &  &  &  &  &  &  &  &  &  &  & 3.40 &  &  & 12.00 &  &  &  &  & $<$ & 0.05 & 0.99 \\ 
  168 & chi2(12)=3.4, p$<$.05 &  &  &  &  &  &  &  &  &  &  &  &  &  &  &  &  &  & 3.40 &  &  & 12.00 &  &  &  &  & $<$ & 0.05 & 0.99 \\ 
  169 & t=.12, p=.34 &  &  &  &  & = & 0.12 &  &  &  &  &  &  &  &  &  &  &  &  &  &  &  &  &  &  &  & = & 0.34 &  \\ 
  170 & p=.12 &  &  &  &  &  &  &  &  &  &  &  &  &  &  &  &  &  &  &  &  &  &  &  &  &  & = & 0.12 &  \\ 
  171 & t(12)=1.2, d=3.4, p=.56 &  &  &  &  & = & 1.20 & 3.40 &  &  &  &  &  &  &  &  &  &  &  &  &  &  & 12.00 &  &  &  & = & 0.56 & 0.25 \\ 
  172 & t(12)=1.2; p=.34 &  &  &  &  & = & 1.20 &  &  &  &  &  &  &  &  &  &  &  &  &  &  &  & 12.00 &  &  &  & = & 0.34 & 0.25 \\ 
  173 & t(12)=1.2 p=.34 &  &  &  &  & = & 1.20 &  &  &  &  &  &  &  &  &  &  &  &  &  &  &  & 12.00 &  &  &  & = & 0.34 & 0.25 \\ 
  174 & t=1.2, df=34, p=.56 &  &  &  &  & = & 1.20 &  &  &  &  &  &  &  &  &  &  &  &  &  &  &  & 34.00 &  &  &  & = & 0.56 & 0.24 \\ 
  175 & t(12)=1.2\verb|^|3, p=4/5 &  &  &  &  & = & 1.73 &  &  &  &  &  &  &  &  &  &  &  &  &  &  &  & 12.00 &  &  &  & = & 0.80 & 0.11 \\ 
  176 & t(12)=1..2, p=.n3 &  &  &  &  & = & 1.00 &  &  &  &  &  &  &  &  &  &  &  &  &  &  &  & 12.00 &  &  &  & = &  & 0.34 \\ 
  177 & t index(12)=1.2, p=.34 &  &  &  &  & = & 1.20 &  &  &  &  &  &  &  &  &  &  &  &  &  &  &  & 12.00 &  &  &  & = & 0.34 & 0.25 \\ 
  178 & r(12)=1.2, p=3.45, R2=6.7 &  &  &  &  &  &  &  & = & 1.20 & = & 6.70 &  &  &  &  &  &  &  &  &  &  & 12.00 &  &  &  & = & 3.45 &  \\ 
  179 & beta=1.2, SE=.34, p$<$.05 &  &  &  &  &  &  &  &  &  &  &  &  &  &  &  &  &  &  &  &  &  &  & 1.20 & 0.34 & 3.53 & $<$ & 0.05 & 0.00 \\ 
  180 & t(12)=1.2, p=5\% &  &  &  &  & = & 1.20 &  &  &  &  &  &  &  &  &  &  &  &  &  &  &  & 12.00 &  &  &  & = & 0.05 & 0.25 \\ 
  181 & t's(12)$>$1.2, p's$>$.05 &  &  &  &  & $>$ & 1.20 &  &  &  &  &  &  &  &  &  &  &  &  &  &  &  & 12.00 &  &  &  & $>$ & 0.05 & 0.25 \\ 
  182 & t(12)$<$=1.2, p$>$=.05 &  &  &  &  & $<$= & 1.20 &  &  &  &  &  &  &  &  &  &  &  &  &  &  &  & 12.00 &  &  &  & $>$= & 0.05 & 0.25 \\ 
  183 & t(12)=1.2, p$<$.05 &  &  &  &  & = & 1.20 &  &  &  &  &  &  &  &  &  &  &  &  &  &  &  & 12.00 &  &  &  & $<$ & 0.05 & 0.25 \\ 
  184 & t(1234)=5.6, p$<$.05 &  &  &  &  & = & 5.60 &  &  &  &  &  &  &  &  &  &  &  &  &  &  &  & 1234.00 &  &  &  & $<$ & 0.05 & 0.00 \\ 
  185 & t(12)=1.2, p$<$.05 &  &  &  &  & = & 1.20 &  &  &  &  &  &  &  &  &  &  &  &  &  &  &  & 12.00 &  &  &  & $<$ & 0.05 & 0.25 \\ 
  186 & t(12)=.34 &  &  &  &  & = & 0.34 &  &  &  &  &  &  &  &  &  &  &  &  &  &  &  & 12.00 &  &  &  &  &  & 0.74 \\ 
  187 & t(12)$<$=$>$1.2, p=.34 &  &  &  &  & $<$=$>$ & 1.20 &  &  &  &  &  &  &  &  &  &  &  &  &  &  &  & 12.00 &  &  &  & = & 0.34 & 0.25 \\ 
   \hline
\end{tabular}
\end{adjustbox}
\end{table}

\end{document}